\documentclass[aps,prb,twocolumn,showpacs,amsmath,amssymb,superscriptaddress]{revtex4-1}
\usepackage{graphicx}% Include figure files
\usepackage{dcolumn}% Align table columns on decimal point
\usepackage{epstopdf}
\usepackage{bm}% bold math
\usepackage{color}
\newcommand{\os}{\onlinecite}

\begin{document}
\title{Toward a realistic description of multilayer black phosphorus: 
From $GW$ approximation to large-scale tight-binding simulations}

\author{A.~N. Rudenko}
\email[]{a.rudenko@science.ru.nl}
\author{Shengjun Yuan}
\author{M.~I. Katsnelson}
\affiliation{Radboud University, Institute for Molecules and Materials, Heijendaalseweg 135, 6525 AJ Nijmegen, The Netherlands}
\date{\today}

\begin{abstract}
We provide a tight-binding model parametrization for black phosphorus (BP) with
an arbitrary number of layers. The model is derived from partially self-consistent
$GW_0$ approach, where the screened Coulomb interaction $W_0$ is calculated
within the random phase approximation on the basis of density functional
theory. We thoroughly validate the model by performing a series of benchmark
calculations,
and determine the limits of its applicability. The application of the model to the 
calculations of electronic and optical properties of multilayer BP demonstrates
good quantitative agreement with \emph{ab initio} results in a wide energy range.
We also show that the proposed model can be easily extended for the case of external fields, 
yielding the results consistent with those obtained from first principles.
The model is expected to be suitable for a variety of realistic problems related to the
electronic properties of multilayer BP including different kinds of disorder, external fields,
and many-body effects.
\end{abstract}


\maketitle

\section{Introduction}
A few-layer black phosphorus (BP) is a novel two-dimensional (2D) semiconductor
with a number of remarkable properties such as strong anisotropy and 
pronounced thickness dependence of its electronic characteristics, which, 
along with high current on-off ratios and
high carrier mobilities, make this material a promising candidate for diverse 
electronic and optical applications [\os{Ling,HLiu,Balendhran,LLi,FXia,Koenig,Buscema}].
Apart from the practical aspect, there is a growing fundamental interest in
BP ranging from attempts to provide insight into the origin of its band 
properties [\os{Appelbaum}] to more exotic and speculative aspects 
including superconductivity [\os{Shao}] and topologically nontrivial 
phases [\os{Zunger}].

From the theoretical perspective, one can distinguish between the two main approaches for 
studying electronic properties in material science. The first one is 
parameter-free first-principles calculations, commonly based on density functional theory (DFT)
and its many-body extensions (e.g., $GW$ approximation). 
Although such methods
generally provide accurate results with respect to the ground state properties, 
their applicability to large systems is very limited due to high
computational cost and poor scalability. At the same time, realistic modeling 
in many cases requires large-scale simulations in order to, for example, 
describe finite-size effects, the presence of interfaces, or different kinds of 
disorder.
In this respect, tight-binding (TB) Hamiltonian models act as an alternative approach 
to the electronic structure problem, providing a way to perform simulations with 
millions of atoms involved. Apart from being computationally more tractable, TB models 
also serve as a playground for exploring rich many-body physics.

Unlike graphene [\os{Katsnelson-Book}], whose electronic 
properties in the low-energy limit are determined by a simple TB
Hamiltonian, involving only one nonequivalent parameter (intersite hopping 
integral, $t$), a reliable theoretical description of a single-layer BP
(known as phosphorene) is considerably more challenging.
A number of low-energy electronic 
properties of pristine phosphorene can be efficiently described in terms of the 
($2 \times 2$) ${\bf k}\cdot {\bf p}$ Hamiltonian [\os{Appelbaum,Rodin,Low1,Low2,Voon,Pereira,Low3}]
with parameters determined to reproduce first-principles calculations.
Being determined in reciprocal space and designed to describe the valence band (VB) and 
conduction band (CB) edges only, the ${\bf k}\cdot {\bf p}$ Hamiltonians are not well suited for
studying real-space problems. Moreover, such models basically rely upon the effective 
mass approximation, whose applicability is not well justified for BP even in the low-energy
range due to the presence of flat bands. Last but not least, although the extension of
the ${\bf k}\cdot {\bf p}$ model appears straightforward to the multilayer case [\os{Low1},\os{Low2},\os{Low3}], it becomes dependent on thickness-dependent parameters, which are \emph{a priori} not known.

Early attempts to provide a real-space model to the electronic structure of BP
were based on molecular orbital theory [\os{Takao}], whose simplified nature and complex orbital
character of BP prevent a quantitatively accurate description [\os{Osada}].
Recently, two of us have proposed a more rigorous real-space model for single- and double-layer BP, 
which was constructed by downfolding the full $G_0W_0$ Hamiltonian to the minimal 
(one interaction site per phosphorus atom) low-energy effective Hamiltonian [\os{Rudenko}].
The latter involves two main parameters of unlike signs corresponding to two nearest-neighbor 
hopping integrals, and a number of less-relevant long-range parameters needed to accurately 
reproduce the quasiparticle VB and CB edges of monolayer and bilayer BP. The model has been 
successfully applied in a number of studies including those related to phosphorene nanoribbons [\os{Ezawa},\os{Sisakht}], 
electric [\os{Ezawa},\os{Dolui}] and magnetic fields [\os{Pereira},\os{Yuan}], different 
kinds of disorder [\os{Yuan}], and realistic modeling of field-effect electronic devices [\os{IEEE}].
However, the applicability of that model is limited to single- and bilayer BP, whereas thicker 
(experimentally available) samples cannot be considered.

In this paper, we report on a revision of the above mentioned model [\os{Rudenko}]. Particularly,
we focus on its modification to describe BP samples with arbitrary 
thickness, ranging from monolayer to bulk. We also improve the quantitative validity of the 
model, which allows us to achieve consistency with experimental results in the bulk limit. 
The proposed model is derived on the basis of accurate 
first-principles calculations within the partially self-consistent 
$GW_0$ approximation and systematically validated by performing a series
of benchmark tests. The model is suitable
for studying large-scale problems and applicable in a 
wide energy range. As a study case, we examine the energy gap dependence 
on the number of layers and also consider the influence of a perpendicular electric field onto 
the electronic structure of BP. Particularly, we study the role of BP thickness 
in the transition from a normal to topological insulator driven by external electric field 
recently predicted for a few-layer BP [\os{Zunger}].

The rest of the paper is organized as follows. In Sec.~II, we start with an overview of previous
first-principles studies of the electronic structure of BP (Sec.~II\,A), provide calculation details (Sec.~II\,B),
and present the results of the $GW_0$ calculations, accompanied by the analysis of
the quasiparticle band structure of a few-layer and bulk BP (Sec.~II\,C). In Sec.~III, we propose the TB model,
describe the parametrization procedure (Sec.~III\,A), and perform a series of calculations in order to assess
its performance (Secs.~III\,B and III\,C). In Sec.~IV, we extend the model by adding an electric field and apply it to multilayer BP.
In Sec.~V, we briefly summarize our results.

\section{Electronic structure of a few-layer BP from first principles}

\subsection{Overview of previous studies}

\begin{figure*}[!t]
\includegraphics[width=0.70\textwidth, angle=0]{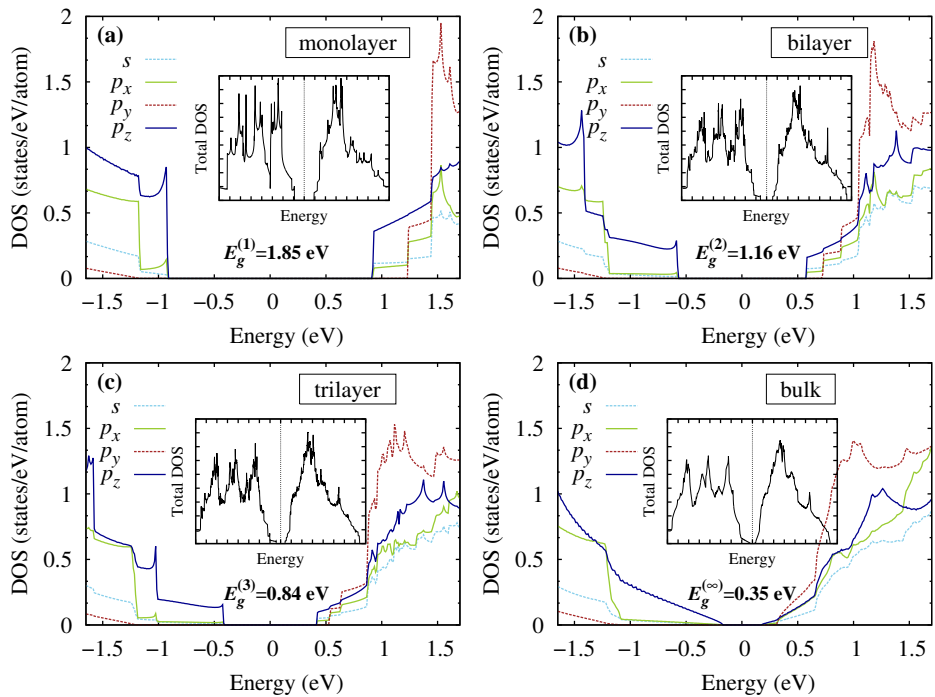}
\caption{(Color online) Orbitally resolved densities of states (DOS) calculated for a few-layer ($n=1$--3) and bulk ($n=\infty$) BP by projecting the $GW_0$ Hamiltonian onto 
the atom-centered Wannier orbitals, corresponding to $s$ and $p_i$ ($i=x,y,z$) symmetries. The total DOS is shown in the inset within the energy range of $-$8 to 8 
eV relative to the band gap center indicated by the vertical dashed line.}
\label{dos}
\end{figure*}

After a few-layer BP became available experimentally, a considerable number of theoretical studies of its band structure 
have been reported. The calculations showed that commonly used DFT in conjunction with local and semilocal exchange-correlation 
approximations does not describe semiconducting properties of bulk BP correctly. Contrary to experimental observations, yielding a narrow gap
of 0.31--0.35 eV [\os{Keyes,Warschauer,Maruyama}] for bulk BP, the local density approximation or generalized gradient approximation (GGA) predict significantly 
smaller or even zero values, depending on a particular computational scheme and lattice parameters [\os{Rudenko},\os{Prytz,Qiao,Liu,Tran,Cai,Kawazoe}]. The utilization
of hybrid functionals [such as Heyd-Scuseria-Ernzerhof (HSE) [\os{hse03},\os{hse06}]], incorporating a nonlocal contribution to the exact exchange, has been shown to partially solve the band gap problem
of bulk BP [\os{Prytz,Qiao,Liu},\os{Cai},\os{Kawazoe}]. However, the performance of such methods depends strongly on a number of empirical parameters determining, for example, 
the screening range and fraction of the exact exchange contribution, which are generally system specific and cannot be systematically determined. This ambiguity results 
in a broad variation of band gaps in a few-layer and bulk BP (see Table \ref{gaps} for an overview).

More consistent results with respect to the band properties can be obtained using the $GW$
approximation [\os{Hedin}], which has been applied to BP in Refs.\,[\os{Rudenko},\os{Tran},\os{Kawazoe},\os{Liang}]. The authors of Ref.\,[\os{Tran}] adopt a non-self-consistent $G_0W_0$ scheme, 
where the screened Coulomb interaction $W_0$ is calculated within the general plasmon pole model [\os{Louie}] and report a band gap of 0.3 eV for bulk BP, which is within
the range of available experimental data. However, the use of a more reliable random phase approximation (RPA) [\os{Shishkin}] within the $G_0W_0$ scheme yields a smaller value of 
0.1 eV [\os{Rudenko}]. More accurate band gap values are supposed to be obtained within the RPA in terms of a partially self-consistent $GW_0$ scheme. Such calculations 
have been recently performed in Ref.\,[\os{Kawazoe}], where the evaluation of $W_0$ was based on hybrid functionals and resulted in significantly higher band gap values for bulk BP (0.58 eV) compared 
to the experimental ones. Therefore, the hybrid functionals do not seem to be an optimal starting point for $GW$ calculations of BP. Physically, this can be attributed to
excessively contracted wave functions, which suppress the screening of the Coulomb repulsion and eventually leads to the band gap overestimation. The closest results to experiment
are obtained by means of the $GW_0$ approach with $W_0$ calculated on top of the GGA wave functions within the RPA (denoted as $GW_0$@GGA thereafter) [\os{Liang}], which yields the gaps of 0.43 and 1.94 eV 
for bulk and monolayer BP, respectively. The latter value is also consistent with recent scanning tunneling spectroscopy measurements of the gap in the spectrum of surface states of cleaved BP 
(2.05 eV) [\os{Liang}].

\subsection{Calculation details}

Here, we first apply the $GW_0$@GGA scheme to calculate the quasiparticle electronic band structures for $n$-layer ($n=1$--3) and bulk BP, which provide reference data
for the subsequent TB model parametrization.
The calculations were performed within the projected 
augmented wave formalism [\os{paw}] as implemented in the \emph{Vienna ab-initio simulation package} ({\sc vasp}) [\os{kresse1996},\os{vasp_paw}]. The Green's functions ($G$)
were first calculated by using the Kohn-Sham eigenvalues and eigenstates and then iterated four times, which proved to be sufficient to achieve numerical convergence [\os{Shishkin2}].
The screened Coulomb interaction ($W_0$) is calculated on the basis of the frequency-dependent dielectric function, $W_0=\epsilon_0^{-1}v$, which, in turn, is computed at the RPA 
level [\os{Shishkin}] as
$\epsilon_0=1-v\chi_0$, where $v$ is the bare Coulomb interaction and $\chi_0$ is the independent particle polarizability. The latter is evaluated by using the DFT-GGA [\os{pbe}] eigenvalues and 
eigenstates 
in the spectral representation. To this end, a numerical integration along the frequency axis containing 70 grid points is performed. In the calculation of the quasiparticle energies, both diagonal
and off-diagonal elements of the self-energy matrix $\Sigma=iGW$ were included. The total energy in the DFT part was converged to within $10^{-8}$ eV. In all calculations, we use an energy cutoff 
of 250 eV for the plane-wave expansion of the wave functions. The number of unoccupied states in $GW$ calculations were set to 90 per atom. In most cases, a {\bf k}-point mesh of ($10 \times 12 \times 1$) 
and ($10 \times 12 \times 4$) was used for the Brillouin zone sampling of a few-layer and bulk BP, respectively. To examine the fine structure of the electronic spectrum of monolayer BP, a denser mesh 
was considered. To obtain smooth band structures, densities of states and optical conductivities, we use an interpolation procedure by making use of the maximally localized Wannier 
functions [\os{Yates,Marzari,wannier90}], which are constructed by projecting the $GW_0$ Hamiltonian onto the entire manifold of the 3$s$ and 3$p$ states of phosphorus. For all the structures, 
we adopt experimental crystal structures of bulk BP [\os{Brown}] and introduce a vacuum layer of $\sim$15\,\AA~in order to minimize spurious effects due to the periodic boundary conditions in slab
calculations. 
The chosen set of parameters ensures that the quasiparticle gaps are accurate to within a few hundredths of eV.
Although some variations in structural parameters have been reported between monolayer and bulk BP [\os{Qiao},\os{Kawazoe}], we intentionally do not consider such effects in our work 
due to the following reasons: (i) to minimize the complexity of the TB model for multilayer BP associated with atomic degrees of freedom, and (ii) to avoid ambiguity in the determination of 
structural parameters for a few-layer BP at the DFT level arising from the variety of different exchange-correlation functionals.

\begin{figure}[tbp]
\includegraphics[width=0.50\textwidth, angle=0]{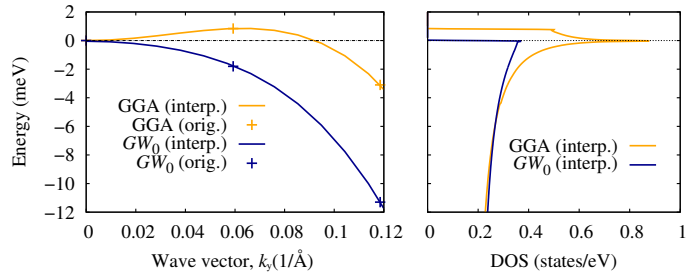}
\caption{(Color online) Left: Fine structure of the VB of monolayer BP calculated along the $\Gamma$-$Y$ direction in the vicinity of the $\Gamma$ point by means of the DFT-GGA and $GW_0$ approaches. 
Points correspond to the original calculations at a ($24 \times 32 \times 1$) {\bf k}-point mesh, whereas lines represent Wannier-interpolated bands. Right: Wannier-interpolated DFT-GGA and $GW_0$ densities of states calculated for the same energy range. Zero energy corresponds to the position of the VB at the $\Gamma$ point.}
\label{subtle}
\end{figure}

\subsection{Quasiparticle electronic properties from partially self-consistent $GW_0$ approximation}

In Fig.~\ref{dos}, we show the densities of states (DOS) calculated for a few-layer ($n=1$--3) and bulk BP within the $GW_0$@GGA scheme. One can see that the calculated value of a band gap of bulk 
BP is 0.35 eV, 
which is within the bounds of experimental variability (0.31--0.35 eV [\os{Keyes,Warschauer,Maruyama}]). Such an agreement justifies the computational approach employed and allows us to expect accurate results for a few-layer BP. Qualitatively,
the $GW_0$ results for a few-layer BP shown in Fig.~\ref{dos} are similar to those reported previously [\os{Rudenko},\os{Qiao},\os{Tran,Cai,Kawazoe}]. In cases of a few-layer BP, DOS exhibits a step like behavior,
which is typical for systems with reduced dimensionality. For the following, it is also worth mentioning that in
all cases considered, the major contribution to the states close to the band gap comes from the $p_z$ states of phosphorus, whereas $p_y$ states have zero contribution at the VB and CB edges.

In the case of monolayer BP, the fine structure of the electronic states near the edge of the VB requires special attention. As has been noticed in previous DFT studies [\os{Rodin}],
the VB maximum is slightly shifted from the $\Gamma$ point in the $\Gamma$-$Y$ direction, which apparently results in an indirect gap in monolayer BP. The deviation of the VB maximum from 
the zone center might result in nontrivial physical properties of BP such as superconducting and ferromagnetic instabilities [\os{Ziletti_VHS}] due to the appearance of the van Hove
singularity close to the VB edge. At the level of the ${\bf k} \cdot {\bf p}$ perturbation theory, a transition from a direct to an indirect band gap in monolayer BP is governed by the magnitudes
of the matrix elements of the momentum operator, corresponding to transitions between the VB and CB [\os{Appelbaum}]. At the same time, well-known inaccuracies of DFT with respect to the
VB and CB positions cannot support the prediction of an indirect gap in monolayer BP. Therefore, it appears appropriate to examine the fine structure of the monolayer VB at the more accurate 
$GW_0$ level. To this end, we perform a comparison between the electronic structures of monolayer BP calculated within the DFT-GGA and $GW_0$ approaches by using a dense 
($24 \times 32 \times 1$) {\bf k}-point mesh. The results are shown in Fig.~\ref{subtle}. We do reproduce the previously reported shift of the VB maximum from the $\Gamma$ point as well as 
the van Hove singularity in DOS calculated at the DFT-GGA level. However, the $GW_0$ results show no indications of such a behavior and support for a \emph{direct} band gap in monolayer BP.

    \begin{table*}[p]
    \centering
    \caption[Bset]{Band gaps (in eV) for monolayer ($n=1$), multilayer ($n=2,3$), and bulk BP ($n=\infty$) calculated at different levels of theory. In the notation of different methods,
$G_0$ and $W_0$ imply that the Green's function and screened Coulomb repulsion in the $GW$ approach are calculated non-self-consistently on the basis of wave functions derived from density 
functional (GGA) or hybrid functional (HSE) calculations, whereas $G$ means a self-consistent calculation of the Green's function. $W'_0$ and $W_0$ denote that the screened Coulomb interaction 
is calculated by using the general plasmon pole model [\os{Louie}] and RPA [\os{Shishkin}], respectively.}
 \begin{tabular}{ccccccccccc}
      \hline
      \hline
& $GW_0$@GGA\footnotemark[1] & TB Model\footnotemark[1] & $GW_0$@GGA\footnotemark[2] & $GW_0$@HSE\footnotemark[3] & $G_0W_0$@GGA\footnotemark[4] & $G_0W'_0$@GGA\footnotemark[5] & HSE\footnotemark[6]                     &                 GGA\footnotemark[7]                            &  Expt.            \\
     \hline
  $n=1$    &  1.85 &  1.84    & 1.94       & 2.41        &  1.60       &  2.00                    &     1.00--1.91                  &     0.80--0.91             &  2.05\footnotemark[8]    \\
  $n=2$    &  1.16 &  1.15    & $\!\!\!\!\!\sim$1.65 & 1.66        &  1.01       &  $\!\!\!\!\!\sim$1.30              &     1.01--1.23                  &     0.45--0.60             &  ---                      \\
  $n=3$    &  0.84 &  0.85    & $\!\!\!\!\!\sim$1.35 & 1.20        &  0.68       &  $\!\!\!\!\!\sim$1.05              &     0.73--0.98                  &     0.20--0.40             &  ---                      \\
$n=\infty$ &  0.35 &  0.40    &  0.43      & 0.58        &  0.10       &  0.30                    &     0.18--0.39                  &     0.00--0.15             &  0.31--0.35\footnotemark[9]    \\
      \hline                  
      \hline
    \label{gaps}
    \end{tabular}
\footnotetext[1]{This work.}
\footnotetext[2]{Reference [\os{Liang}].}
\footnotetext[3]{Reference [\os{Kawazoe}].}
\footnotetext[4]{Reference [\os{Rudenko}].}
\footnotetext[5]{Reference [\os{Tran}].}
\footnotetext[6]{References [\os{Prytz,Qiao,Liu,Cai,Kawazoe}].}
\footnotetext[7]{References [\os{Prytz,Qiao,Liu,Rudenko,Tran,Cai,Kawazoe,Rodin}].}
\footnotetext[8]{This value corresponds to a gap in the spectrum of surface states of bulk BP (Ref.\,[\os{Liang}]).}
\footnotetext[9]{References [\os{Keyes,Warschauer,Maruyama}].}
    \end{table*}

\section{Tight-binding model for multilayer BP and its validation}

\begin{figure}[tbp]
\includegraphics[width=0.48\textwidth, angle=0]{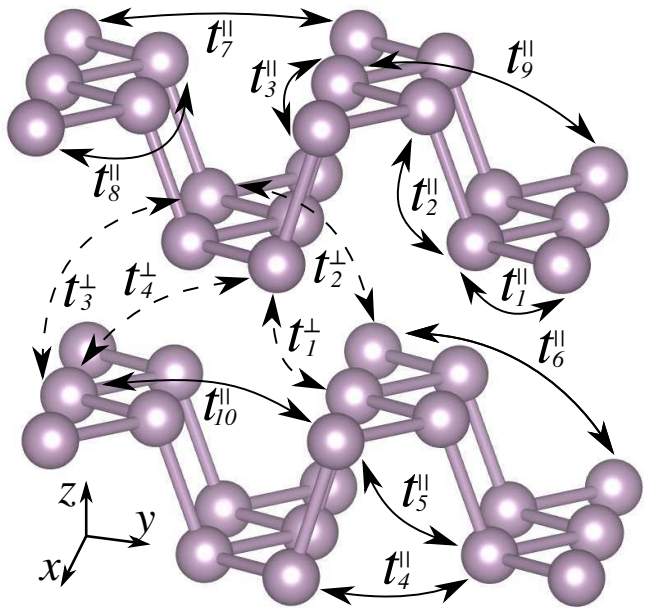}
\caption{(Color online) Schematic representation of the hopping parameters for the TB model [Eq.~(\ref{tb_hamilt})] parametrized in this work for multilayer BP. The corresponding values are given
in Table \ref{hoppings_table}.}
\label{hoppings}
\end{figure}

\subsection{Parametrization procedure}

The effective TB model considered in this work is given by the effective four-band Hamiltonian, describing one electron per lattice site,
\begin{equation}
H=\sum_{i\neq j} t^{\parallel}_{ij} c_i^{\dag}c_j + \sum_{i \neq j} t^{\perp}_{ij}c_i^{\dag}c_j,
\label{tb_hamilt}
\end{equation}
where $i$ and $j$ run over the lattice sites, $t^{\parallel}_{ij}$ ($t^{\perp}_{ij}$) is the intralayer (interlayer) hopping parameter between the $i$ and $j$ sites, and $c_i^{\dag}$ ($c_j$) is the
creation (annihilation) operator of electrons at site $i$ ($j$). We note that in contrast to the model used in our previous works [\os{Rudenko},\os{Yuan}], the Hamiltonian given by Eq.~(\ref{tb_hamilt}) does
not contain on-site terms, meaning that electrons at all sites have equivalent energies even for multilayer BP.

To parametrize the model given by Eq.~(\ref{tb_hamilt}), we use a procedure similar to Ref.\,[\os{Rudenko}], which is as follows. We first map the
entire manifold of valence and conduction states of phosphorus monolayer onto the subspace of effective $p_z$-like orbitals (four orbitals per unit cell) 
being relevant for the low-energy part of the VB and CB. 
To this end, we first use the original Bloch states $|\psi_{n{\bf k}}\rangle$ obtained from the $GW_0$ calculations and
construct a new subspace of Bloch-like states $| \tilde{\psi}_{n{\bf k}} \rangle$,
\begin{equation}
|\tilde{\psi}_{n{\bf k}}\rangle = \sum_{m=1}^P U_{mn}^{\bf k} |\psi_{m{\bf k}}\rangle,
\end{equation}
where $P$ is the total number of states included into the $GW_0$ calculations and $U_{mn}^{\bf k}$ is a rectangular matrix obtained by projecting the $|p_z\rangle$ states onto 
the Bloch states $|\psi_{n{\bf k}}\rangle$ and using the disentanglement procedure proposed in Ref.\,[\os{Souza}]. Having obtained $|\tilde{\psi}_{n{\bf k}}\rangle$, we construct
an effective ($4 \times 4$) Hamiltonian in reciprocal space $\tilde{H}_{mn}({\bf k})$, which is achieved by performing a unitary transformation of the original $GW_0$ Hamiltonian $H_{mn}^{\bf k}$ in 
the Bloch subspace. The resulting reciprocal-space Hamiltonian
\begin{equation}
\tilde{H}_{mn}^{\bf k}=\langle \tilde{\psi}_{m{\bf k}}|H^{\bf k}| \tilde{\psi}_{n{\bf k}} \rangle
\end{equation}
is then transformed into the real space, $H_{mn}^{\bf R}=\sum_{\bf k}e^{-i{\bf k}\cdot {\bf R}}\tilde{H}_{mn}^{\bf k}$. The resulting real-space Hamiltonian $H_{mn}^{\bf R}$ is determined in the 
basis of Wannier functions $| w_n^{\bf R} \rangle=\sum_{{\bf k}}e^{-i{\bf k}\cdot {\bf R}}|\tilde{\psi}_{n{\bf k}}\rangle$, corresponding to the $p_z$-like orbitals.

Despite low dimensionality of $H_{mn}^{\bf R}$, its matrix elements (hopping parameters) decay slowly with distance, resulting in a large number of small parameters. In order to make the resulting model 
more
tractable, we ignore the parameters beyond the cutoff radius of $\sim$5.5\,\AA, which are typically smaller than 0.01 eV. To restore the quality of the truncated Hamiltonian, we reoptimize the 
remaining parameters in such a way that 
they provide an accurate description of the band structure in the low-energy region. To this end, we minimize the following least squares functional, 
$F(\{t_i\})=\sum_{n,{\bf k}}[\varepsilon_{n,{\bf k}}^{GW_0}(\{t_i\})^2 - \varepsilon_{n,{\bf k}}^{\mathrm{TB}}(\{t_i\})^2]$, where $\{t_i\}$ are hopping parameters and
$\varepsilon_{n,{\bf k}}^{GW_0}$ ($\varepsilon_{n,{\bf k}}^{\mathrm{TB}}$) is an eigenvalue of the corresponding ($GW_0$ or TB model) Hamiltonian $H^{\bf k}(\{t_i\})$. $n$ and {\bf k} are the band index
and momentum vector, respectively, which run over the relevant region in the vicinity of the band gap. In the case of monolayer, this region involves the valence and conduction bands only. To parametrize
the TB Hamiltonian for bilayer, we adopt a similar strategy. In this case, we introduce interlayer hopping parameters, while the intralayer parameters remain fixed. Also, we take into account the 
splitting of the valence and conduction bands upon the optimization of the hoppings, which is crucially important for the applicability of the model to multilayer BP.
The obtained set of intralayer (\{$t^{\parallel}_i\}$) and interlayer (\{$t^{\perp}_i\}$) hoppings are then applied without any corrections to BP with a larger number of layers.

\subsection{Electronic structure}

The resulting hopping parameters are schematically shown in Fig.~\ref{hoppings} and listed in Table \ref{hoppings_table}. Overall, our model involves ten intralayer and four interlayer hoppings. As has been
previously noticed [\os{Rudenko},\os{Ezawa}], the main features of the band structure of monolayer BP can be qualitatively described by only two largest hopping parameters ($t_1^{\parallel}$ and $t_2^{\parallel}$). 
The band gap at the $\Gamma$ point is determined in this case by a simple expression, $E^{(1)}_g(\Gamma)\approx2|t_2^{\parallel}|-4|t_1^{\parallel}|$. For bilayer BP, the degeneracy of the VB and CB is lifted
if a nearest-neighbor interlayer hopping ($t_1^{\perp}$) is introduced. This results in a reduction of the band gap, given now by
$E^{(2)}_g(\Gamma)\approx2|t_2^{\parallel}|\sqrt{1+(t_1^{\perp}/t_2^{\parallel})^2}-4|t_1^{\parallel}|-2|t_1^{\perp}|$.
In order to quantitatively reproduce the quasiparticle spectrum of BP including accurate {\bf k} dependence of the VB and CB as well as their splitting in the multilayer case, 
a larger number of hopping parameters is required.

\begin{figure*}[tbp]
\includegraphics[width=0.70\textwidth, angle=0]{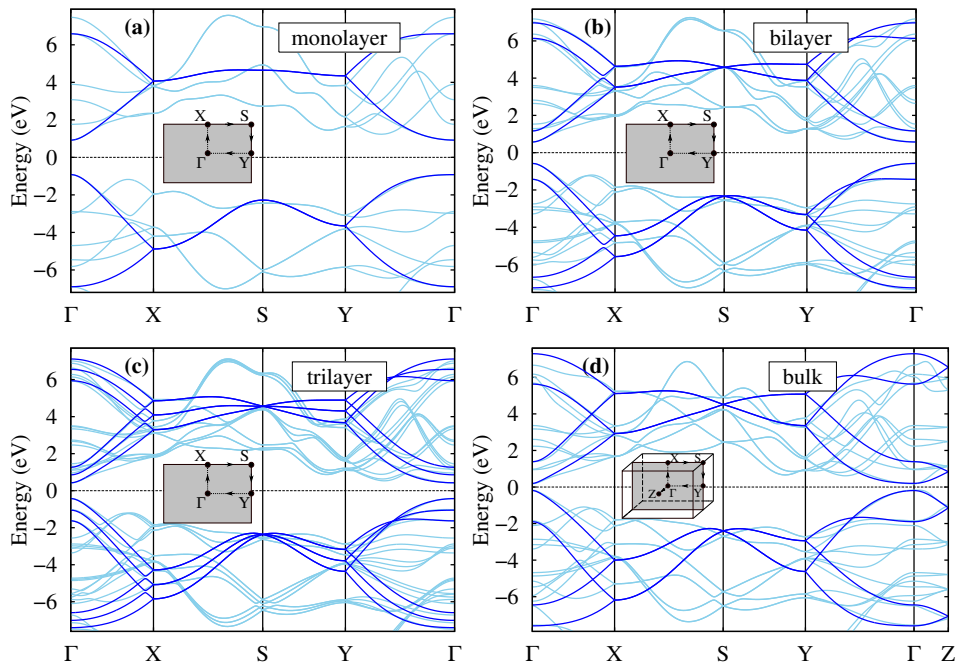}
\caption{(Color online) Electronic band structures calculated for a few-layer ($n=1$--3) and bulk BP by using the $GW_0$ approximation (light lines) and within the TB model given 
by Eq.~(\ref{tb_hamilt}) with parameters listed in Table \ref{hoppings_table} (dark lines). Zero energy corresponds to the center of the gap. High-symmetry points of the Brillouin zone are shown
in the insets.
}
\label{bands}
\end{figure*}

In Fig.~\ref{bands}, we show the band structures calculated within the derived TB model in comparison with the full bands obtained from $GW_0$ calculations.
One can see that the TB model accurately describes the results of $GW_0$ calculations in the low-energy region not only for monolayer and bilayer BP, but also for trilayer and bulk structures. 
Since the band properties of trilayer and bulk BP have not been used as a reference during the model parametrization, it is natural to expect
the applicability of the presented model to BP with an arbitrary number of layers.

    \begin{table*}[!bt]
    \centering
    \caption[Bset]{Intralayer ($t^{\parallel}$) and interlayer ($t^{\perp}$) hopping parameters (in eV) obtained in terms of the TB Hamiltonian [Eq.~(\ref{tb_hamilt})] for multilayer BP. $d$ and
          $N_c$ denote the distances between the corresponding interacting lattice sites and coordination numbers for the given distance, respectively. The hoppings are schematically shown
in Fig.~\ref{hoppings}.}
 \begin{tabular}{cccccccccccccc}
      \hline
      \hline
 &  \multicolumn{3}{c}{Intralayer} & & \multicolumn{3}{c}{Intralayer} & & & \multicolumn{3}{c}{Interlayer} \\
\cline{2-4}
\cline{6-8}
\cline{11-13}
No.& \, $t^{\parallel}$ (eV)\,& \, $d$ (\AA) \, & \, $N_c$ \, & No.& \, $t^{\parallel}$ (eV)\,& \, $d$ (\AA) \,& \, $N_c$ \,&     & No. & \, $t^{\perp}$ (eV)\,& \, $d$ (\AA) \, & \, $N_c$ \,\\
     \hline
  1& \   $-$1.486  \    &  \    2.22  \  &  \   2   \  & 6  &    \ \,\,\, 0.186  \   &   \  4.23 \   &      1     &     &  1  &    \  \,\,\,\,\,0.524  \      & \    3.60   \  & \  2 \     \\
  2& \   \,\,\, 3.729  \    &  \    2.24  \  &  \   1   \  & 7  &    \ $-$0.063  \   &   \  4.37 \   &      2     &     &  2  &    \  \,\,\,\,\,0.180  \      & \    3.81   \  & \  2 \     \\
  3& \   $-$0.252  \    &  \    3.31  \  &  \   2   \  & 8  &    \ \,\,\, 0.101  \   &   \  5.18 \   &      2     &     &  3  &    \ $-$0.123  \      & \    5.05   \  & \  4 \     \\
  4& \   $-$0.071  \    &  \    3.34  \  &  \   2   \  & 9  &    \ $-$0.042  \   &   \  5.37 \   &      2     &     &  4  &    \ $-$0.168  \      & \    5.08   \  & \  2 \     \\
  5& \   $-$0.019  \    &  \    3.47  \  &  \   4   \  & 10 &    \ \,\,\, 0.073  \   &   \  5.49 \   &      4     &     &  5  &    \  \,\,\,\,\,0.000  \      & \    5.44   \  & \  1 \     \\
      \hline                  
      \hline
    \end{tabular}
\label{hoppings_table}
    \end{table*}

\begin{figure}[!tbp]
\includegraphics[width=0.45\textwidth, angle=0]{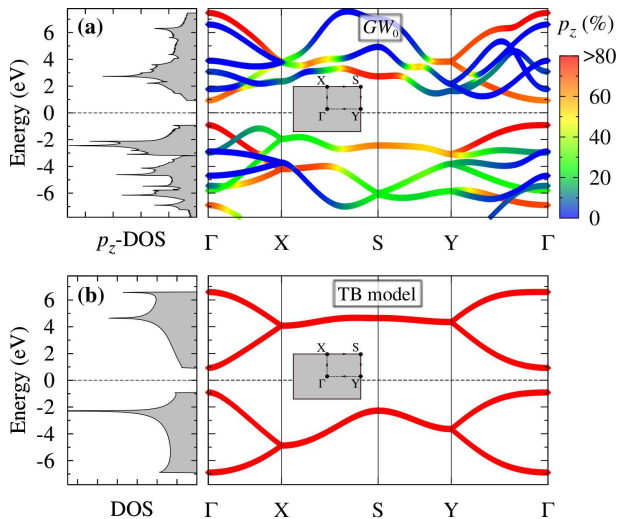}
\caption{(Color online) Band structure and density of states (DOS) calculated for monolayer BP by using (a) $GW_0$ approach and (b) TB model of this work. In (a), 
the DOS are projected onto the $p_z$ states, whereas in the band structure their contribution is shown by color. High-symmetry points of the Brillouin zone are shown in the insets.}
\label{bands_proj}
\end{figure}

To explicitly demonstrate that the obtained TB Hamiltonian is represented in a physically meaningful orbital subspace corresponding to the $p_z$-like states, 
we consider the case of monolayer BP, for which we project the full $GW_0$ band structure onto the $p_z$ states [see Fig.~\ref{bands_proj}(a)] and compare 
it with the model bands [Fig.~\ref{bands_proj}(b)]. From the projected $GW_0$ bands shown in Fig.~\ref{bands_proj}(a) one can clearly recognize four distinct bands
having predominantly $p_z$ symmetry, whose contribution is shown by color. By comparing those with Fig.~\ref{bands_proj}(b) it becomes evident that the model
provides an effective representation of the $p_z$-like states. As can be inferred from Fig.~\ref{bands_proj}(a) and will be shown below, the states of 
the other symmetries do not contribute to direct interband transitions within an energy range of up to several eV, which basically determines the limits of the applicability
of the presented TB model.

\begin{figure}[!tbp]
\includegraphics[width=0.50\textwidth, angle=0]{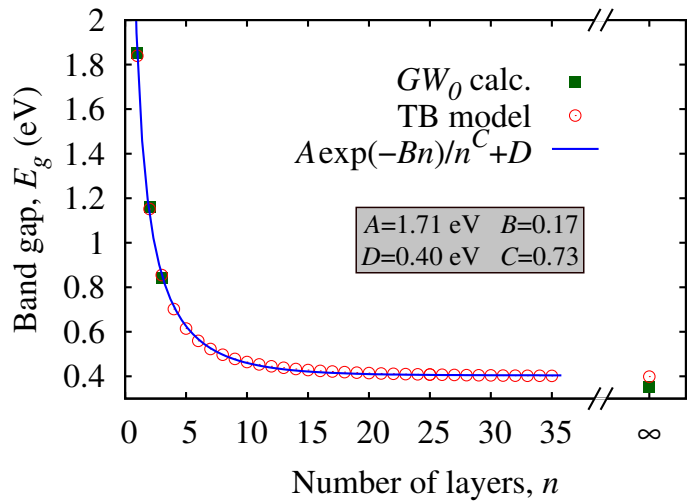}
\caption{(Color online) Layer dependence of a band gap in BP
calculated by using the $GW_0$ approximation, TB model presented in this work 
and by an empirical expression $E_g^{(n)}=A\,\mathrm{exp}(-nB)/n^C+D$ with parameters
$A$, $B$, $C$, and $D$ fitted to the TB model.}
\label{gap}
\end{figure}

Having obtained a TB model applicable for multilayer BP, it is instructive
to analyze the the 
band gap dependence on the number of layers. In Fig.~\ref{gap}, we show
the corresponding dependence calculated within the TB model, which can be accurately
fitted by the expression $E_g^{(n)}=A\,\mathrm{exp}(-nB)/n^C+D$ with parameters $A$, $B$, $C$, and $D$
given in the inset of Fig.~\ref{gap}. One can see that along with a power law decay, being important at
small $n$, there is a pronounced exponential decay, becoming dominant at large $n$. 
Our result is thus different from the previously proposed power law expected from 
a simple quantum confinement picture [\os{Tran}].

\subsection{Optical properties}

To further validate our model, we calculate the frequency-dependent
optical conductivity $\sigma_{\alpha \beta}(\omega)$ calculated for the undoped case by means of the $GW_0$ approach 
and TB model for $n=1$--3 layer and bulk BP. Within the $GW_0$, we evaluate $\sigma_{\alpha \beta}(\omega)$ through the Brillouin zone integration
using the following form of the Kubo-Greenwood formula in the independent-particle approximation [\os{Allen}]
\begin{equation}
\sigma_{\alpha \beta}(\omega)=\frac{i\hbar}{N_k\Omega}\sum_{\bf k}\sum_{mn}\frac{f_{m{\bf k}}-f_{n{\bf k}}}{\varepsilon_{m{\bf k}}-\varepsilon_{n{\bf k}}}
 \frac{ \langle n{\bf k}|j_{\alpha}|m{\bf k}\rangle  \langle m{\bf k}|j_{\beta}|n{\bf k}\rangle } {\varepsilon_{m{\bf k}}-\varepsilon_{n{\bf k}}-(\hbar \omega +i\eta)},
\label{Kubo}
\end{equation}
where $\Omega$ is the unit cell area, $N_k$ is the number of ${\bf k}$ points used for the Brillouin zone sampling, $|m{\bf k}\rangle$ is the Wannier-interpolated Bloch state [\os{Yates}], corresponding to the 
$m$th eigenvalue $\varepsilon_{m{\bf k}}$ of the $GW_0$ Hamiltonian $H^{GW_0}_{\bf k}$, $f_{n{\bf k}}=\mathrm{exp}(\beta \varepsilon_{n{\bf k}}+1)^{-1}$ is the Fermi-Dirac occupation factor 
involving the inverse temperature $\beta$, $j_{\alpha}$ is the $\alpha$ component of the current operator, and $\eta$ is a smearing parameter. The Brillouin zone was sampled 
by $\sim$10$^7$ and $10^8$ {\bf k} points for 2D (a few layer) and 3D (bulk) calculations, respectively. To demonstrate the advantage of the derived TB model for studying realistic samples, 
we apply the TB Hamiltonian [Eq.~(\ref{tb_hamilt})] to calculate $\sigma_{\alpha \beta}(\omega)$ for a few-layer ($n=1$--3) and many-layer ($n$=100) BP in real space. 
To this end, we use the tight-binding propagation method [\os{Yuan},\os{Yuan2010}], in which $\sigma_{\alpha \beta}(\omega)$ is calculated conceptually similar to Eq.~(\ref{Kubo}) but
by considering explicit evolution of the current operator in time [e.g., see Eq.~(30) of Ref.\,[\os{Yuan2010}]] instead of diagonalization of large matrices. 
The sample size was taken to contain $\sim$10$^7$ atoms in each case considered with periodic boundary conditions applied in lateral ($xy$) directions.
In both methods, we restrict ourselves to the diagonal components of $\sigma_{\alpha \beta}(\omega)$ only.
We stress that $\sigma_{\alpha\beta}(\omega)$ is calculated within a single-particle approximation, meaning that the excitonic effects are neglected. Such effects are
proven to be relevant for a reliable description of the optical spectra of monolayer and a few-layer BP, but they become insignificant in the bulk limit [\os{Tran}].

\begin{figure}[!tbp]
\includegraphics[width=0.51\textwidth, angle=0]{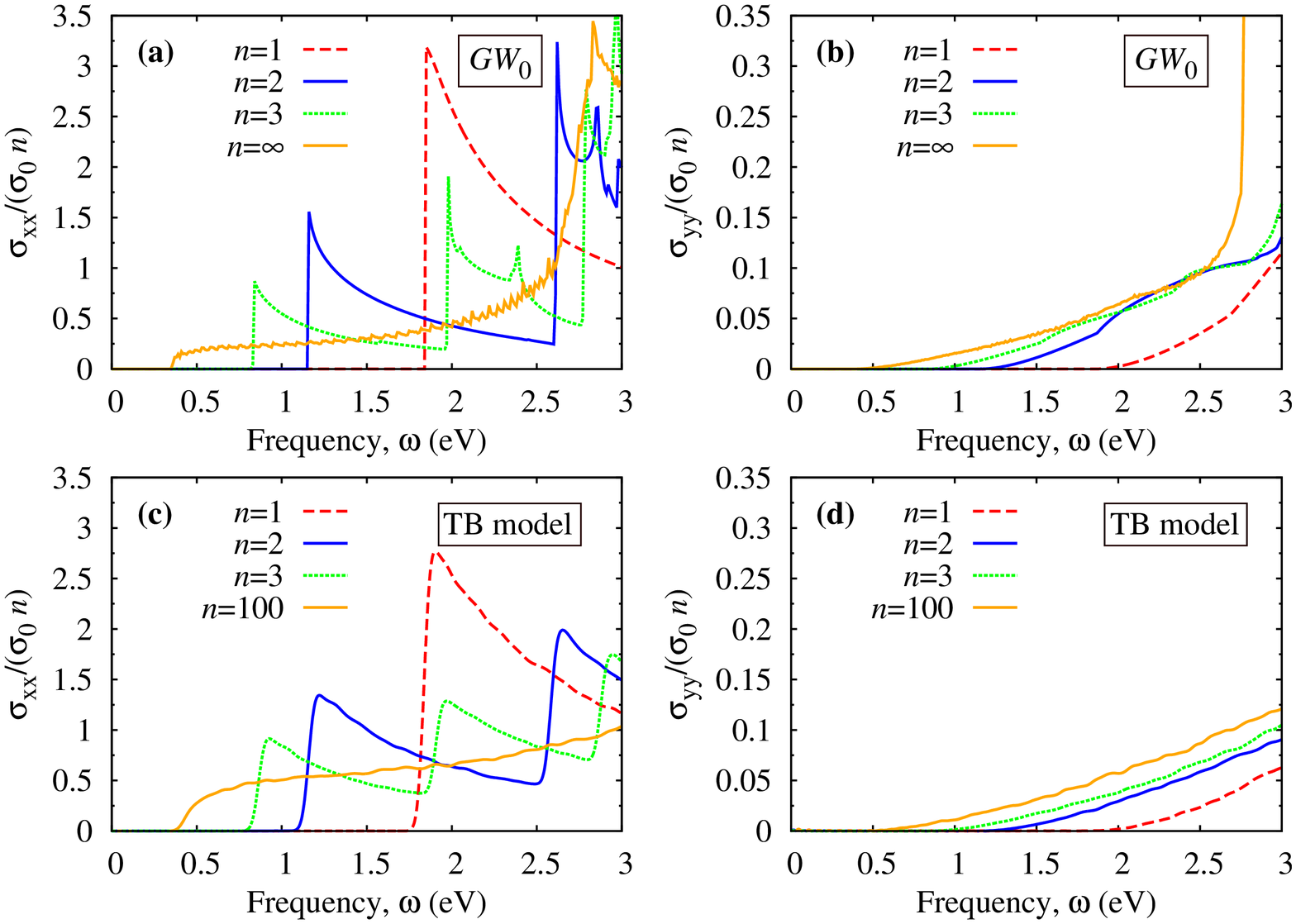}
\caption{(Color online) Optical conductivities for a few-layer ($n=1$--3) and bulk BP calculated 
along the armchair ($\sigma_{xx}$) and zigzag ($\sigma_{yy}$) directions using the Kubo formula [Eq.~(\ref{Kubo})] on 
the basis of the $GW_0$ approach and TB model presented in this work. $\sigma_{xx(yy)}$ are given 
per layer in terms of the universal optical conductivity of graphene ($\sigma_0=e^2/4\hbar$). Within
the TB model, bulk BP is approximated by 2D BP with a large number of layers ($n=$100). In all
cases, we set the temperature to 300 K.}
\label{cond}
\end{figure}

The results of our calculations are shown in Fig.~\ref{cond}. In line with previous studies [\os{Low1},\os{Yuan}], we observe
strong anisotropy between the conductivities in different directions and well-pronounced peaks along the armchair direction
of a few-layer BP, associated with the discrete character of the band structure close to the VB and CB
edges. As can be seen from Fig.~\ref{cond}, the optical conductivities obtained with the use of the TB model are in very good 
agreement with the results of $GW_0$ calculations in a wide frequency range up to 2.0 eV.
The agreement in the range of 2.0--2.5 eV can be considered as satisfactory but it is becoming worse for structures with a large number of
layers. At larger frequencies ($\omega>2.5$ eV), the TB model still shows reasonable agreement with the $GW_0$ results for a few-layer BP, 
but becomes apparently inapplicable to many-layer systems (including bulk), which is due to the decreased gap, allowing for transitions 
between the states not included in the construction of the TB model.
Despite being relatively close to the band gap, those states do not contribute to the optical conductivity at the lower frequencies since the 
expression for $\sigma_{\alpha\beta}(\omega)$ [Eq.~(\ref{Kubo})] involves only direct transitions between the VB and CB.
We note that for transport and optical properties involving \emph{indirect} 
transitions between the VB and CB (e.g., in scattering processes) a reliable frequency range for the TB model will be more limited and determined entirely 
by the consistency between the quasiparticle and model bands shown in Fig.~\ref{bands}.

\section{Effect of electric field on the band structure of multilayer BP}

We now consider an extension of our model to the case of an electric field $E_z$ perpendicular to the surface. For simplicity,
we restrict ourselves to bilayer BP, for which the extended Hamiltonian reads

\begin{equation}
H=H_0^{(2)}+eE_zz,
\label{extension}
\end{equation}
where the first term in the right-hand side corresponds to the unperturbed Hamiltonian for bilayer given by 
Eq.~(\ref{tb_hamilt}) and the second term plays the role of a layer-dependent on-site potential.
We note that in what follows, we consider an \emph{unscreened} electric field only, that is, we neglect explicit treatment of 
polarization and local-field effects. In other words, $E_z$ can be regarded as a local electric field assumed to be
constant inside the sample. $E_z$ can be related to real external electric field $E_z^{\mathrm{ext}}$ upon taking into account 
thickness-dependent transverse dielectric permittivity $\varepsilon_z(d)$ and finite-size effects. In a first approximation, one can 
take $E_z^{\mathrm{ext}}=\varepsilon_z E_z$, where $\varepsilon_z$ is the transverse dielectric permittivity of bulk BP 
($\varepsilon_z \sim 8.3$ [\os{Nagahama}]).

\begin{figure}[tbp]
\includegraphics[width=0.48\textwidth, angle=0]{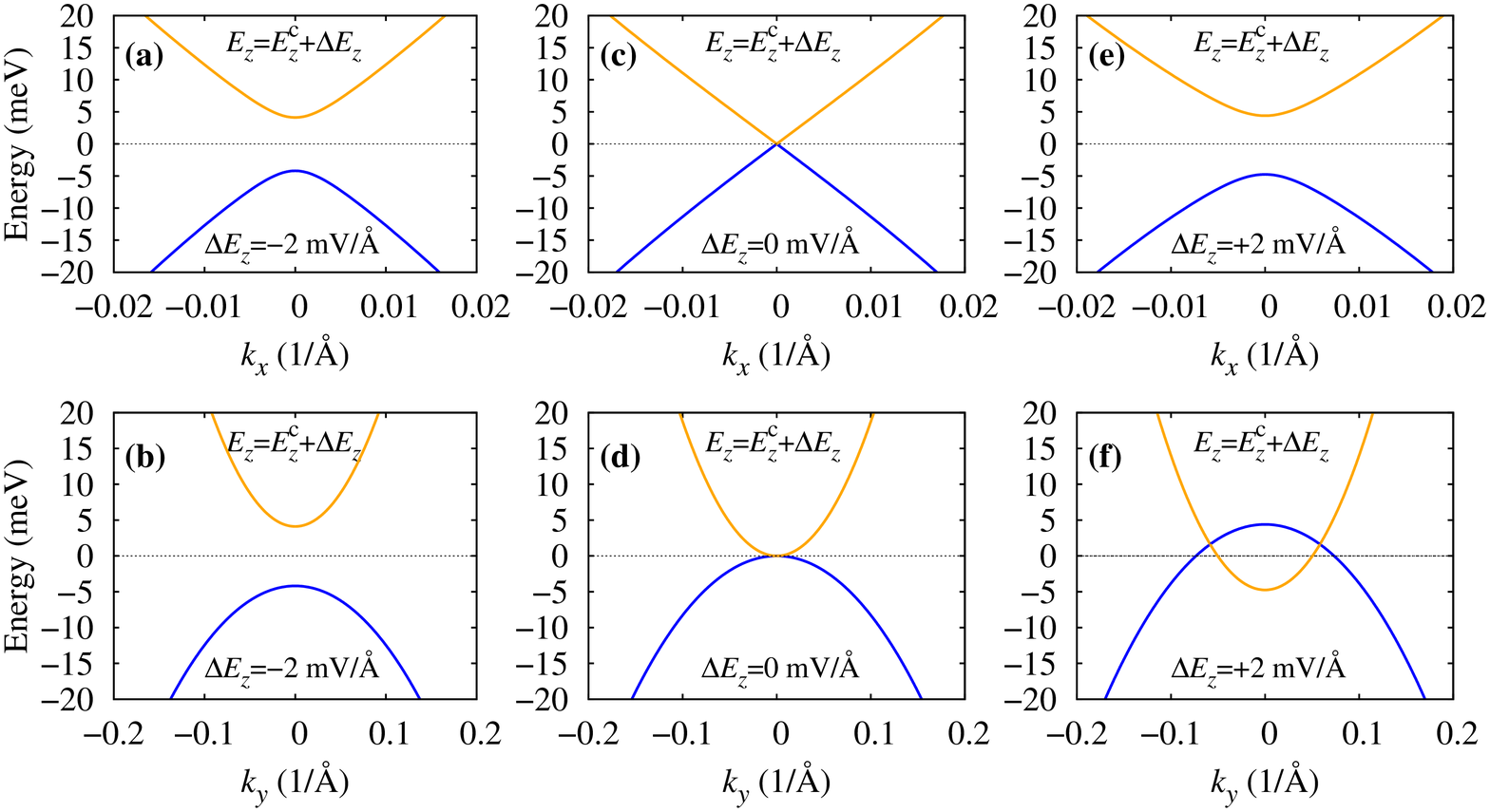}
\caption{(Color online) Band structures of bilayer BP calculated in the vicinity of the $\Gamma$ point for different magnitudes 
of the electric field $E_z=E_z^\mathrm{c}+\Delta E_z$, where $E_z^\mathrm{c}=341$ mV/\AA~is a critical field at which 
the band gap closes, and $\Delta E_z$ takes the values of $-2$, 0, and +2 mV/\AA. Top (bottom) panels correspond to the 
bands calculated along the $X$-$\Gamma$-$X$ ($Y$-$\Gamma$-$Y$) directions. Valence and conduction bands are indicated by blue and orange, respectively. 
Zero energy corresponds to the center of the gap at the $\Gamma$ point.}
\label{biased}
\end{figure}

In Fig.~\ref{biased}, we show the low-energy part of the band structure calculated for three
representative electric fields. In the presence of an electric field, the electronic bands shift due to
the difference of the interlayer potential which is a manifestation of the Stark effect. From Fig.~\ref{biased} 
one can see that the VB and CB shift in different directions toward the band gap center. This causes 
a decrease of the band gap with increasing field, which reaches zero at $E_z=341$ mV/\AA.
At higher field the band inversion is observed, as can be seen from Fig.~\ref{biased}(f). Our results 
obtained using the TB model are thus consistent with previous DFT calculations for a few-layer BP and phosphorene 
nanoribbons [\os{Zunger},\os{Dolui},\os{HGuo,YanLi,QWu}].

It is interesting to note the existence of a Dirac-like linear dispersion along the armchair direction ($X$-$\Gamma$-$X$) at the
critical electric field, $E^\mathrm{c}_z$ [Fig.~\ref{biased}(c)], which appears around the $\Gamma$ point. A qualitatively different
situation is observed in the zigzag direction ($Y$-$\Gamma$-$Y$), where the dispersion turns out to be quadratic [Fig.~\ref{biased}(d)]. 
At higher fields ($E_z>E_z^\mathrm{c}$) the Dirac point disappears in the armchair direction, whereas two band crossings
appear along the zigzag direction as a result of the band inversion.

\begin{figure}[t]
\includegraphics[width=0.48\textwidth, angle=0]{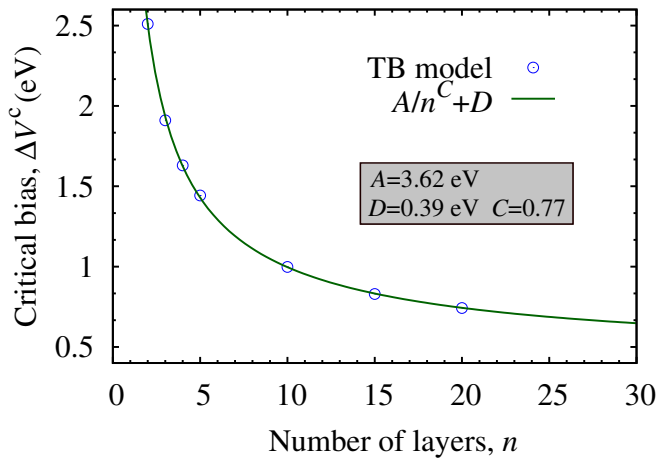}
\caption{(Color online) Critical bias potential $\Delta V^\mathrm{c}(n)$ required to close the band gap in $n$-layer BP plotted as a function of the number of layers.
The bias potential is applied within the TB model of this work [see Eq.~(\ref{extension})] between the top and bottom planes of the corresponding BP samples neglecting 
the screening effects. Points correspond to the TB calculations, whereas lines correspond to the fitting via $\Delta V^\mathrm{c}(n)=A/n^C+D$.}
\label{criticalV}
\end{figure}

Finally, we calculate the evolution of the critical bias potential,
$\Delta V^\mathrm{c}=eE_z^\mathrm{c}d$ with the number of BP layers $n$, which is applied between the top and bottom planes of an
$n$-layer sample separated by the distance $d$.
In Fig.~\ref{criticalV}, the corresponding dependence is shown. Since $\Delta V^\mathrm{c}$ is related to the original band gap at zero field,
it is natural to expect that the same form of the functional dependence as in Fig.~\ref{gap} (i.e., a power law with exponential cutoff) can be 
used to parametrize $\Delta V^\mathrm{c}(n)$. We find, however, that in the present case the prefactor $B$ in the argument of the exponential is significantly smaller
($B<0.01$). This allows us to fit the critical bias potential as $\Delta V^\mathrm{c}(n)=A/n^C+D$, where $A$, $C$, $D$ are fitting parameters given in the inset of Fig.~\ref{criticalV}. 
We conclude, therefore, that $\Delta V^\mathrm{c}(n)$ (Fig.~\ref{criticalV}) exhibit a significantly weaker dependence on the number of BP layers than 
the band gap, $E_g^{(n)}$ (Fig.~\ref{gap}).

\quad

\section{Conclusions}

We have proposed an effective TB model for multilayer BP with arbitrary
thickness, which is parametrized on the basis of partially self-consistent
$GW_0$ approximation. The model shows good performance with respect to
static band properties as well as transport characteristics of multilayer
BP compared to the $GW_0$ results. 
In contrast to previously proposed ${\bf k}\cdot {\bf p}$ Hamiltonians 
for BP, our model (i) directly applicable in real space; (ii) goes beyond the effective 
mass approximation; and (iii) accurately reproduces low-energy electronic 
properties of multilayer BP without the need for additional scaling parameters. 
On the other hand, the proposed model is substantially less computationally 
demanding than any first-principles calculations, which makes calculations 
with millions of atoms possible.
This allows us to expect its suitability 
for use in investigations of a wide range of phenomena, particularly in
large-scale simulations of realistic BP (e.g., with disorder or in the 
presence of external fields) and as a starting point for studying many-body
effects in BP. As an example of the model extension, we considered the case
of an electric field applied to multilayer BP, which allowed us to determine
the thickness dependence of the critical bias potential required to reach the regime
of the band inversion previously predicted by first-principles 
calculations. We also found that the critical bias potential decays significantly slowly
with the number of BP layers than does the corresponding band gap.

\section{Acknowledgments}

This research has received funding from the European Union Seventh 
Framework Programme under Grant Agreement No.~604391 Graphene Flagship and
from the European Research Council Advanced Grant program (Contract No.~338957).
Computational time provided by the Netherlands National Computing Facilities
(NCF) is acknowledged.

\end{document}